\documentstyle[preprint,aps]{revtex}

\begin{document}
\draft
\preprint{CALT-68-2010}

\title{Optical scattering angles and Bose-stimulated motion \\
in cold atomic gas}
\author{H. David Politzer}
\address{California Institute of Technology, Pasadena, California 91125\\
{\tt politzer@theory.caltech.edu}}
\date{October 6, 1995}
\maketitle

\begin{abstract}
Bose statistics imply a substantial enhancement at small angles for light
scattering off a cold, Bose gas. The enhancement increases dramatically at
the Bose-Einstein temperature. This phenomenon could be utilized to
eliminate almost entirely the heating of the gas by a weak probe light beam.
\end{abstract}



\narrowtext

A salient feature of Bose statistics is that the transition probability to a
particular final state is proportional to $n_f+1$, where $n_f$ is the number
of bosons hitherto in that final state. For photons, this is the origin of
stimulated emission due to the presence of an external electromagnetic
field. The analog for the motion of bosonic atoms is the enhancement of
scattering into already occupied states. This is the essential ingredient in
the dynamics of the formation of a Bose-Einstein condensate as the
temperature drops below the critical value, $T_{\text{c}}.$ A boson
transition probability is also linear in the number of identical particles
in the initial state. The present note points out how these factors effect
simple photon--atom scattering in the Born approximation (e.g. for weak
light). At low temperatures, there is substantial enhancement for small
photon scattering angles (corresponding to small momentum transfers) from
stimulated scattering of atoms into already occupied low momentum states.
Below $T_{\text{c}}$, this enhancement dramatically shifts to even smaller
angles due to the contribution of transitions from occupied atomic momentum $%
{\bf p\neq 0}$ states to the Bose condensate ${\bf p=0}$ state and {\it vice
versa}. Aside from seeing Bose statistics in action, this may have a
practical implication for the optical investigation of cold atoms. The
incident light intensity --- and consequent total heating of the gas --- can
be reduced by several orders of magnitude for a given, desired signal
strength by putting the detector at very small angle. This may allow
non-destructive measurements on super-cold gases.

I treat here the simplest case of a uniform medium and study the photon
scattering rate per unit time and volume into a given solid angle. A further
simplifying approximation is that the atoms are treated as an ideal, i.e.
non-interacting, Bose gas, so that the multiparticle states are described by
occupancies of single particle states. Whether appropriate in detail to a
particular atom trap depends on trap shape and size, atomic density, and
beam wavelength and size. However, the same qualitative features will emerge 
\cite{java}.

The underlying process is just Rayleigh scattering (or perhaps on resonance
to enhance the rate). I focus on the angular dependence of the rate for
fixed incident wavelength. With very sub-Doppler temperatures $T$ (and with
a very narrow light frequency band if using a resonance --- narrow compared
to the atomic line width), there is no relevant energy dependence in the
scattering amplitude even though we sum over possible initial and final
atomic kinetic energies. This is because the range of energies involved in
the thermal sum over atoms is negligible compared to the energy scale of
variation of the scattering amplitude. The only angular dependence in the
underlying ``unstimulated'' rate comes from the photon polarizations: $\cos
^2\theta $ for polarizations in the scattering plane, $1$ for perpendicular, 
$\frac 12(1+\cos ^2\theta )$ for unpolarized --- all of which I treat as $1$
for small angles. (The polarization difference may have some potential
interest because it vanishes as $\theta \rightarrow 0$ and, hence, weights
angles differently.)

For a given, pure, many-particle quantum state with $n_i$ atoms in an
initial one-particle state and $n_f$ atoms initially in the one-particle
state into which one of the $n_i$ is scattered, the scattering rate is
proportional to $n_i(n_f+1)$. The thermally averaged rate is, therefore,
proportional to $\langle n_i(n_f+1)\rangle $, where the brackets denote the
thermal expectation. For $i\neq f$, which is guaranteed by having $\theta
\neq 0$, this is equal to $\langle n_i\rangle (\langle n_f\rangle +1)$ if we
ignore interactions between the atoms themselves. Hence, the needed thermal
information is simply the thermal average occupancies of the single particle
states $i$ and $f$. It is simplest to evaluate these in terms of a chemical
potential, which is then adjusted to produce the desired mean total number
of particles. Note, however, that for bosons at very low temperatures, if
any state has a large occupancy, the expected fluctuations in the occupancy
are comparably large. The actual, physical situation of interest will
determine which description (fixed particle number or fixed chemical
potential) is more convenient.

It is useful to divide the scattering rate into three terms. (a) The total
``unstimulated'' rate, i.e., the $1$ term in $n_f+1$, is normalized to $%
N_{total}\cdot 1$, where $N_{total}$ is the total target atom density and
where the Rayleigh rate is normalized to $1$. (b) The scattering from $p\neq
0$ occupied states to $p\neq 0$ occupied states is represented by a double
integral over two Bose thermal occupation factors ($\cdot 1$). And (c) below 
$T_c$, there is scattering from $p\neq 0$ occupied states to the $p=0$
state, with occupation $N_o(\tau )$, and {\it vice versa;} these two pieces
give identical contributions. Because the integrals over $p$'s converge as $%
p\rightarrow 0$, they do not account for a macroscopic occupancy of the $p=0$
state, which must be treated separately. Of course, there is no $N_o\cdot
N_o $ term at non-zero angle. Note that the ``unstimulated'' rate is linear
in the density, while the stimulated terms are quadratic.

Choose units such that $\hbar =c=1$. The convenient third choice of units is 
$mc^2k_BT_c=1$, where $m$ is the atom mass. Let ${\bf \Delta }$ be the
dimensionless measure of the momentum transfer to the photon, i.e. expressed
in the natural units of the problem. Then 
\begin{equation}
\Delta =\left| {\bf \Delta }\right| =\frac{\sin \theta }{\cos \frac \theta 2}%
\frac{h\nu }{\sqrt{mc^2k_BT_c}}\simeq \theta \cdot k  \eqnum{1}
\end{equation}
where $\theta $ is the photon scattering angle, $k$ is the magnitude of its
momentum in these units, and $\nu $ is its frequency. For laser light
scattering off alkali atoms condensing at temperatures of order $10^{-7}K$, $%
\theta $ is of order 0.1 to 0.01 times $\Delta $. (If the conversion factor
between $\theta $ and $\Delta $ were much smaller, e.g. with yet lower atom
density and, hence, lower $T_c$, the phenomena discussed below might be
limited to inaccessibly small angles.)

Define a scaled temperature $\tau \equiv T/T_c$.\- Then $R(\Delta ,\tau )$,
the total scattering rate normalized to the Rayleigh rate off a density of $%
N_{total}$, is given by 
\begin{eqnarray}
R(\Delta ,\tau ) &=&1  \eqnum{2a} \\
&&+N_{total}^{-1}\int \frac{d^3{\bf p}d^3{\bf p}^{\prime }\delta ^3({\bf p-p}%
^{\prime }-{\bf \Delta })}{[\lambda ^{-1}(\tau )e^{p^2/2\tau }-1][\lambda
^{-1}(\tau )e^{p^{\prime 2}/2\tau }-1]}  \eqnum{2b} \\
&&+2\cdot \frac{N_o(\tau )}{N_{total}}\int \frac{d^3{\bf p}d^3{\bf p}%
^{\prime }\delta ^3({\bf p}^{\prime })\delta ^3({\bf p-p}^{\prime }{\bf %
-\Delta })}{[\lambda ^{-1}(\tau )e^{p^2/2\tau }-1]}  \eqnum{2c}
\end{eqnarray}

where 
\begin{eqnarray*}
N_{total} &=&\int d^3{\bf p}(e^{p^2/2}-1)^{-1}=\frac 12(2\pi )^{3/2}\zeta (%
\frac 32) \\
N_o(\tau ) &=&(1-\tau ^{3/2})N_{total}\Theta (1-\tau )
\end{eqnarray*}
\begin{eqnarray*}
\lambda (\tau ) &\equiv &e^{\mu (\tau )/\tau } \\
&=&1-\{\frac 9{16\pi }[\zeta (\frac 32)]^2(\tau -1)^2+{\cal O}((\tau
-1)^3)\}\Theta (\tau -1)
\end{eqnarray*}
and where $\mu (\tau )$ is the chemical potential and the Riemann $\zeta
(3/2)\simeq 2.61$. The coefficient in $\lambda (\tau )$ of $(\tau -1)$ above 
$\tau =1$ is determined by requiring $N_{total}$ (which in these units is
the critical density) to be independent of $\tau $. The virtue of the
dimensionless variables $\Delta $, $\tau $, and $N_{total}$ is that a single
function $R(\Delta ,\tau )$ describes all physical situations, and
interesting phenomena occur when the variables are roughly ${\cal O(}1).$
The actual light frequency, atom density, and atom mass then determine the
translation of $(\Delta ,\tau )$ into $(\theta ,T)$.

The integral in term (2b) cannot be evaluated in closed form. After using
the $\delta $-function and integrating over angles, one is faced with a
one-dimensional integral, depending on $\Delta $ and $\tau $, that must be
performed numerically. Before addressing the results, note that, despite the 
$\Theta $-functions in $N_o$ and $\lambda $, both $R(\Delta ,\tau )$ and $%
\partial R/\partial \tau $ are continuous functions of $\tau $. (The latter
is not immediately obvious but can be confirmed by an explicit evaluation of
the contribution to $\partial R/\partial \tau $ from the $\tau $-dependence
of $\lambda (\tau )$ at $\tau =1$; this can be done analytically. The
discontinuity in this contribution to the $\tau $ derivative is exactly
cancelled by the $\tau $-dependence of term (2c). The $\tau $-dependence
arising from the $e^{p^2/\tau }$'s is clearly continuous.)

The Bose final state enhancement of the total scattering, i.e., integrated
over all angles, is a factor of $N_{total}$, independent of $\tau .$ What
changes with $\tau $ is the angular dependence of that enhancement. As $\tau 
$ decreases, it is concentrated at smaller and smaller angles --- because
the thermal distribution is at lower and lower momenta. What happens as we
go below $\tau =1$ is that the finite fraction of the Bose enhancement
involving the $p=0$ Bose-Einstein condensate has a different angular
distribution, concentrated at somewhat smaller angles. It is the singular
nature of the Bose distribution as ${\bf p}\rightarrow 0$ and $\tau
\rightarrow 1^{+}$ that keeps $R$ and $\partial R/\partial \tau $ formally
continuous. Nevertheless, the numerical evaluation reveals a dramatic
crossover for sufficiently small $\Delta $.

The most interesting ranges of parameters are displayed in figure (1). Above
the critical temperature, $\Delta $'s of order $1$ or below show a Bose
enhancement (solid lines) of a factor of two to five. Below the critical
temperature, for $\Delta $'s like $0.1$ or below, the enhancement can jump
to several hundred or thousand with only a ten to twenty percent drop in
temperature. The dashed curves are the contribution to the normalized rate
coming from scatterings involving the $p=0$ states, either as targets or
final states. These scatterings account for a substantial fraction but by no
means all of the enhancement. At fixed $\Delta $, as $T\rightarrow 0$, the
enhancement must ultimately disappear. (It is then concentrated into yet
smaller angles).

The above effects are really just simple kinematics that follow from the
Bose occupation numbers. However, they have a potential practical
application in cold atom optics. By collecting scattered light at small
angles, one could significantly reduce the incident intensity needed to
produce a given signal strength. The thermal average of the enhanced small
angle scatterings do not heat the gas at all. They collectively remove
precisely as much energy as they add because they involve all possible
transitions from the thermal distribution back into the thermal
distribution. The isotropic component (2a) is always present, and it does
always contribute to heating. However, if the rate at some milliradians is $%
500$ times greater than at $90$ degrees, then one could reduce the incident
intensity and, hence, total heating rate by $500$ by moving the detector to
small angles and still have the same signal strength.

The present discussion makes the simplifying assumption that the target gas
is uniform over the beam size. One practical effect of finite sample size is
that diffraction limits the smallest meaningful angles. For current
experiments\cite{corn}, the conversion of $\Delta $ to a laboratory
scattering angle puts the interesting angles at the border of what might be
observed. A small atom trap may also produce significant variation in the
position space support of the wave functions. This can be used, of course,
to advantage in identifying the ground state occupancy \cite{corn}. The
formulae presented here are more appropriate to a trap that is large
compared to the beam size or more like an ideal box. Nevertheless, the
behavior for small, harmonic traps should be similar \cite{java}. There are
again three classes of states: unoccupied high energy states, a thermal
occupation distribution of low energy excited states, and, finally, the
ground state, with a significant occupancy. And the ``unstimulated'' rate is
linear in the density while the stimulated rate is quadratic. One salient
difference between trapped atoms and a uniform medium is that, with an
external potential, momentum is not conserved \cite{java}. In an infinite,
uniform medium, the super-stimulated ground-state-to-ground-state
transitions correspond to zero photon scattering angle. The corresponding
transitions in a trap are analogs of the M\"{o}ssbauer effect and are
dominant for photon momentum transfers up to about double the largest
typical ground state atom momentum. It is only beyond the corresponding
maximum scattering angle that trap ground-state-to-excited-state and
excited-state-to-excited-state scattering are the largest effects. Note,
however, that if the light scattering is performed after a significantly
long time interval of free expansion, i.e. after suddenly turning off the
trapping fields \cite{corn}, the Bose-stimulated enhancement will be
confined to increasinly smaller angles. This is because, under free 
propagation, different momentum components become spatially separated. 
So the overlap between the scattered atom, of a particular final momentum, and
previously existing atoms with that same momentum decreases rapidly with
time.

Many recent theoretical discussions of optical effects in low temperature
Bose gases have focused on far subtler issues, such as coherence in forward
scattering at low \cite{dali} or high \cite{shlyap,hdp} densities (relative
to the optical wavelength) or with pulsed light \cite{lew&you}, or the
effects of finite trap size \cite{java} or transition line shapes \cite{you}%
. (For a comprehensive review, see \cite{lewetal}.) Although the fact that
stimulated scattering occurs at small angles has often been mentioned in
passing, a quantitative estimate of the magnitudes, angles, and sharpness
with temperature of the effect had not been given. As emphasized in this
context by Javanainen \cite{java2}, to manifest any dramatic consequence of
Bose statistics, the situation must involve a substantial number of
identical particles that can be in the same place. For the present
discussion to be relevant, the figure of merit is the number of atoms in the
volume defined by the inverse of the momentum transfer.

\acknowledgements 
The author thanks K. Libbrecht and J. Preskill for several very helpful
observations and suggestions. This work was supported in part by the U.S.
Dept. of Energy under Grant No. DE-FG03-92-ER40701.

\begin{figure}[tbp]
\caption{The scattered light, total enhancement factor, $R(\Delta ,T/T_c)$,
including Bose stimulated atomic motion (solid lines) versus{\it \ }%
temperature at different reduced angles $\Delta $ (see eqn.\ (1)). The
dashed curves are the condensate contributions (i.e., from eq. (2c)).}
\end{figure}


\begin{references}
\bibitem{java}  J. Javanainen, Phys. Rev. Lett. {\bf 72}, 2375 (1993).

\bibitem{corn}  M.H. Anderson, J.R. Ensher, M.R. Matthews, C.E. Wieman, and
E.A. Cornell, Science {\bf 269}, 198 (1995).

\bibitem{dali}  O. Morice, Y. Castin, and J. Dalibard, Phys. Rev. A {\bf 51}%
, 3896 (1995).

\bibitem{shlyap}  B.V. Svistunov and G.V. Shlyapnikov, Sov. Phys. JETP {\bf %
70}, 460 (1990); {\bf 71}, 71 (1990).

\bibitem{hdp}  H. D. Politzer, Phys. Rev. A {\bf 43}, 6444 (1991).

\bibitem{lew&you}  M. Lewenstein and L. You, Phys. Rev. Lett. {\bf 71}, 1339
(1993).

\bibitem{you}  L. You, M. Lewenstein, and J. Cooper, Phys. Rev. A {\bf 50},
R3565 (1994).

\bibitem{lewetal}  M. Lewenstein, L. You, J. Cooper, and K. Burnett, Phys.
Rev. A {\bf 50}, 2207 (1994).

\bibitem{java2}  J. Javanainen, unpublished.
\end{references}
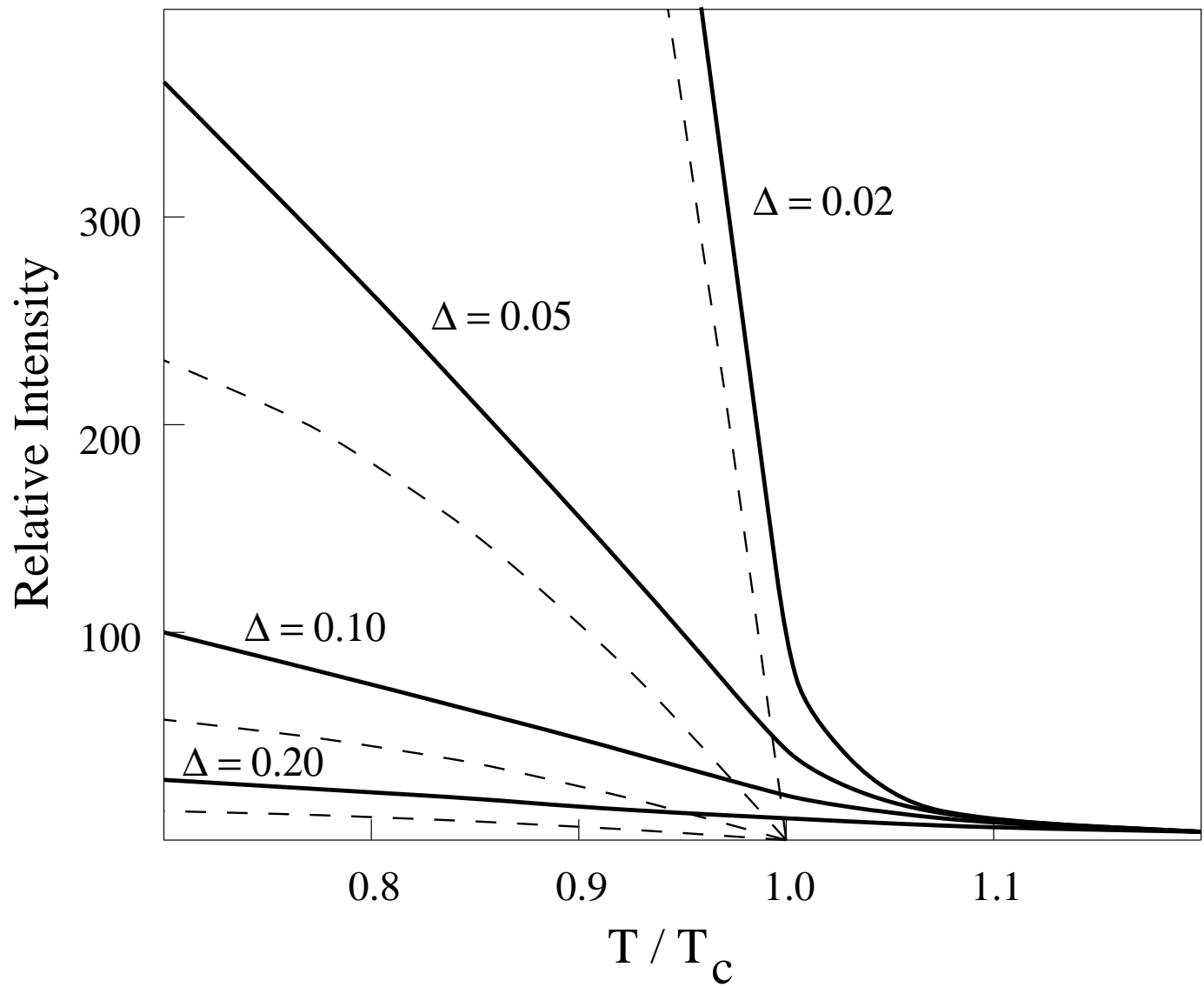
\end{document}